\documentclass[twoside]{ilcws07}
\usepackage[latin1]{inputenc}
\usepackage[dvips]{graphicx,epsfig,color}
\usepackage{wrapfig,rotating}
\usepackage{amssymb,amsmath,array}

\begin{document}
\title{Calorimeter Assisted Tracking Algorithm for SiD}
\author{Dmitry Onoprienko$^1$ and Eckhard von Toerne$^1,^2$
\vspace{.3cm}\\
1- Kansas State University - Physics Department \\
116 Cardwell Hall Manhattan, KS 66506  - USA
\vspace{.1cm}\\
2- Bonn University - Institute of Physics \\
Nussallee 12, D-53115 Bonn - Germany\\
}

\maketitle

\begin{abstract}
Calorimeter-assisted track finding algorithm takes advantage of the finely 
segmented electromagnetic calorimeter proposed for the SiD detector 
concept by looking 
for "MIP stubs" produced by charged particles in the calorimeter, and using 
them as seeds for pattern recognition in the tracker. The algorithm allows 
for efficient reconstruction of tracks that cannot be found using seeds 
provided by the vertex detector, even if standalone pattern recognition in the outer tracker is 
difficult. The algorithm has been implemented as a package in the org.lcsim 
framework. Current status of the package and its performance in
non-prompt tracks reconstruction are described.

\end{abstract}

\section{Introduction}

The development of the calorimeter assisted tracking algorithm was originally 
motivated by the need to reconstruct long-lived particles in the SiD detector~\cite{sid}.

To minimize multiple scattering and energy loss in the central tracker 
while providing excellent vertex finding capabilities and high precision 
measurement of the charged tracks momenta, 
the SiD baseline design utilizes a compact five-layer 
silicon pixel vertex detector and a five-layer silicon strip outer tracker, 
possibly with no stereo and limited Z-segmentation in the barrel. 

The standard track finding algorithm developed for the SiD relies on identifying 
tracks in the vertex detector, where pattern recognition is simplified by 
the fact that precise three-dimensional information  is available for each 
hit. The tracks are then propagated into the outer tracker, 
additional hits are picked up, and the track curvature is measured. 
This algorithm has been 
demonstrated~\cite{Sinev:2005xw} to achieve high efficiency in reconstructing most types of tracks. 
However, its heavy reliance on seeds provided by the vertex detector 
raises a number of questions that need to be addressed. One obvious issue is
that the tracks that originate outside the third layer of the vertex detector cannot
be reconstructed using this approach, since they do no leave enough hits in
pixels to generate a seed. Decay products of $K_{S}^{0}$ and $\Lambda$ are important examples
of particles producing such non-prompt tracks. 
The detector should also be capable of
detecting new physics signatures that would include long-lived exotic particles 
like those predicted by some gauge-mediated supersymmetry breaking scenarios.

In order to address this issue, a track finding algorithm has been developed 
that uses electromagnetic calorimeter to provide seeds for pattern recognition 
in the tracker. Fine segmentation of the EM calorimeter allows for detection 
of traces left by minimum ionizing particles - so called MIP stubs - 
and using them to determine the track entry point, direction, and sometimes 
curvature with a precision sufficient for extrapolating the track back 
into the tracker.

\section{Algorithm}

With calorimeter assisted tracking, track finding goes through the following main steps:

\begin{enumerate}

\item The standard vertex detector seeded track finder is run. Tracker hits
that are associated with successfully reconstructed tracks are removed.

\item MIP stubs in the electromagnetic calorimeter are identified. Several alternative
algorithms can be used at this step, such as a generic nearest-neighbor clustering
followed by user-controlled cluster quality cuts,
or a dedicated MIP stub finder developed specifically for the SiD geometry.
For each MIP stub, a seed track is created, and initial helix parameters are determined.

\item Seed tracks are
extrapolated back into the tracker, picking up hits in each layer. Every time a hit
is added to the track, the track is re-fitted to get a more precise estimate of its
parameters. If multiple hit candidates are found in a given layer, the track finding
process is branched and several independent track candidates are created.

\item Quality cuts are applied to track candidates; duplicate tracks that share too many
hits are discarded.

\item Vertex finder is run; if track intersections are found, the original
particles that produced these secondary vertices are reconstructed.

\end{enumerate}

The algorithm has been implemented as a package (\texttt{org.lcsim.recon.cat}) 
in the Java based org.lcsim
framework~\cite{lcsim}. The package design is highly modular,
allowing easy substitution of components implementing different algorithms
at various stages of the event processing. Most of the track finder parameters can
be set at run time, making it possible for the user to tune the algorithm and 
use it as a part of more complex reconstruction strategies that can involve
multiple track finders and multiple passes through the data.

The code can be run both standalone and inside JAS3 interactive shell~\cite{Johnson:2001di}.
The implementation is largely decoupled from any particular geometry, 
which allows the algorithm to be used for studying and optimizing 
a wide range of SiD detector 
options, as well as other ILC detectors designs.

\section{Performance}

\begin{figure}
\begin{minipage}[t]{0.5\linewidth} 
\centering
\includegraphics[width=1\linewidth]{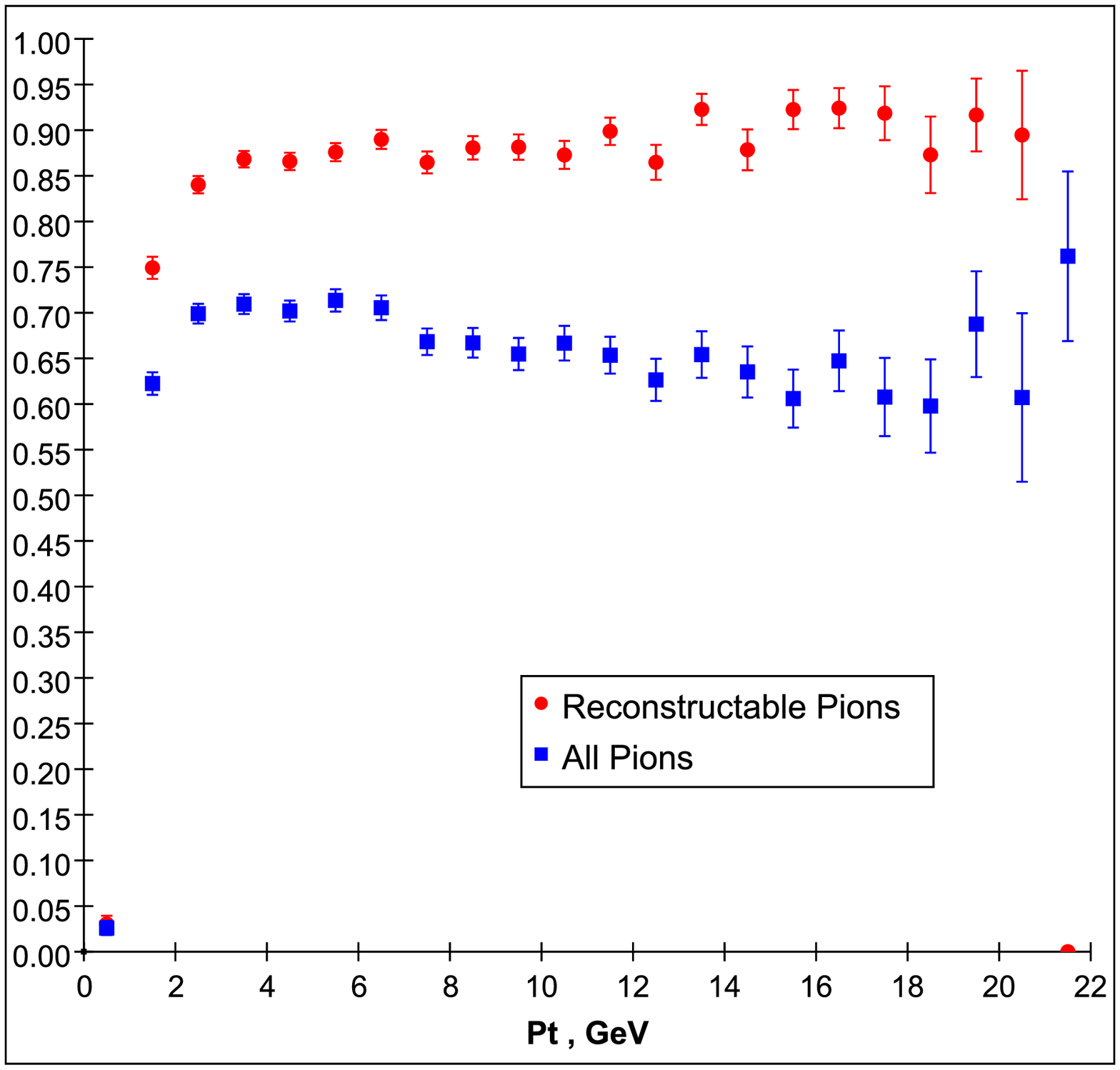}
\caption{Reconstruction efficiency for charged $\pi$ mesons from $K_{S}^{0}$ decays 
in single $K_{S}^{0}$ events, as a function of transverse momentum.}
\label{Fig:1}
\end{minipage}
\hspace{0.5cm} 
\begin{minipage}[t]{0.5\linewidth}
\centering
\includegraphics[width=1\linewidth]{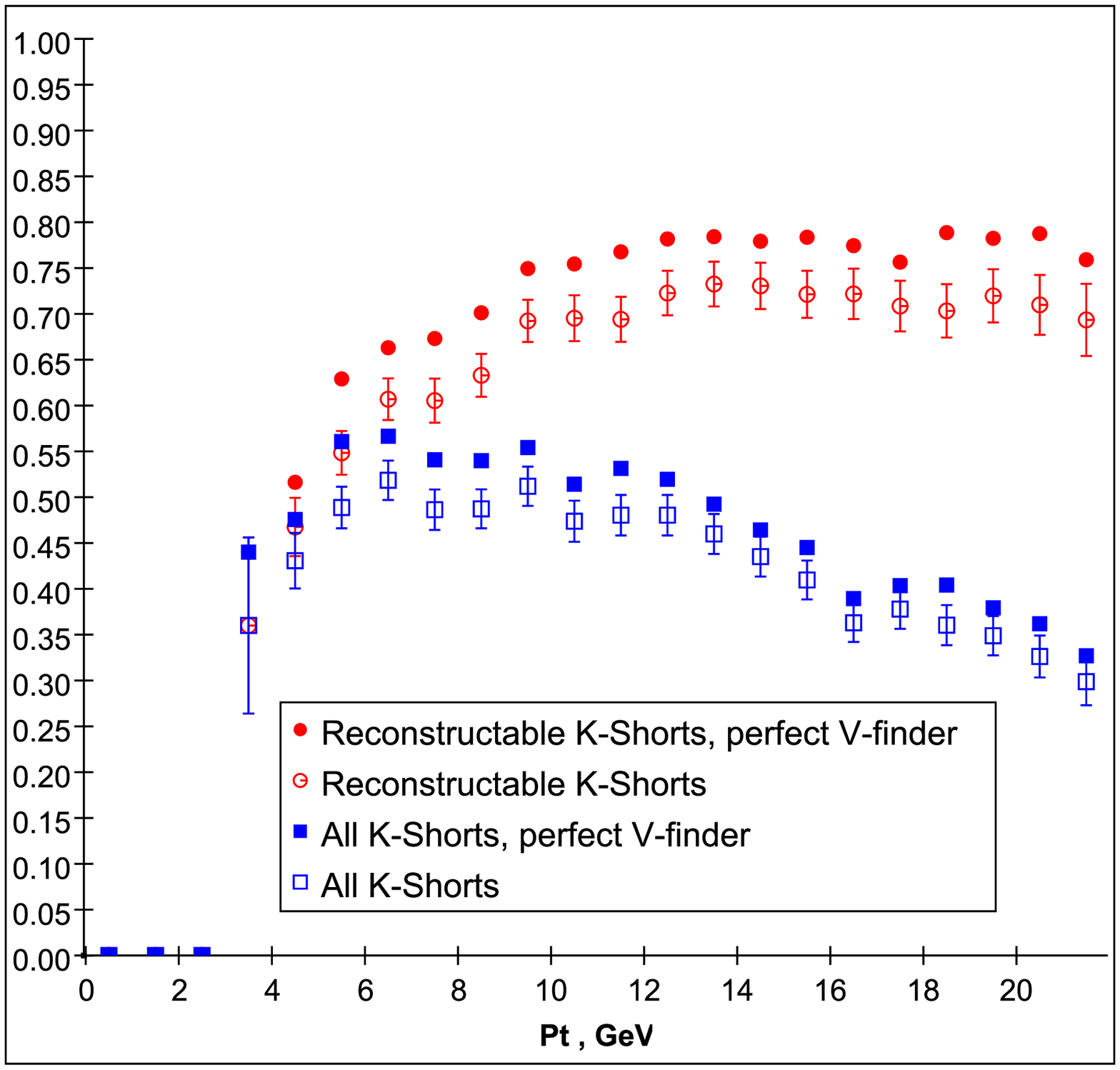}
\caption{$K_{S}^{0}$ reconstruction efficiency as a function of transverse momentum. Single $K_{S}^{0}$ events.}
\label{Fig:2}
\end{minipage}
\end{figure}

The package was tested by reconstructing simulated events in the SiD detector.
For this study, the \texttt{"SiD00"} version of the detector design was used. Since 
proper charge deposition and digitization code was not yet available, 
point-like hits produced by the simulation software were smeared with 
the expected position resolution. In the outer
tracker barrel, no stereo layers and 10~cm silicon strips parallel to the beam
line were modeled. Strips on opposite sides of the endcap disks were assumed to
be perpendicular to each other, forming 90~degrees stereo superlayers, 
with 10~cm strip length.

Figure \ref{Fig:1} shows reconstruction efficiency obtained with the calorimeter
assisted tracking algorithm for charged $\pi$ mesons produced in
single $K_{S}^{0}$ events, as a function of transverse momentum. Pions are considered 
reconstructable if they leave hits in at least 3
tracker layers. Figure \ref{Fig:2} shows reconstruction efficiency for $K_{S}^{0}$
in the same type of events.
Hollow points refer to the actual efficiency obtained with the current version
of the package. Since the currently used vertex finder is a simplistic tool 
expected to be replaced by a more advanced algorithm once the latter is ported to the 
org.lcsim framework, the efficiency that would be obtained with a perfect
vertex finder is also shown (filled points). 
$K_{S}^{0}$ is considered reconstructable if it
decays in the sensitive area of the detector (inside the third layer of the outer
tracker).

\begin{figure}
\begin{minipage}[t]{0.5\linewidth} 
\centering
\includegraphics[width=1\linewidth]{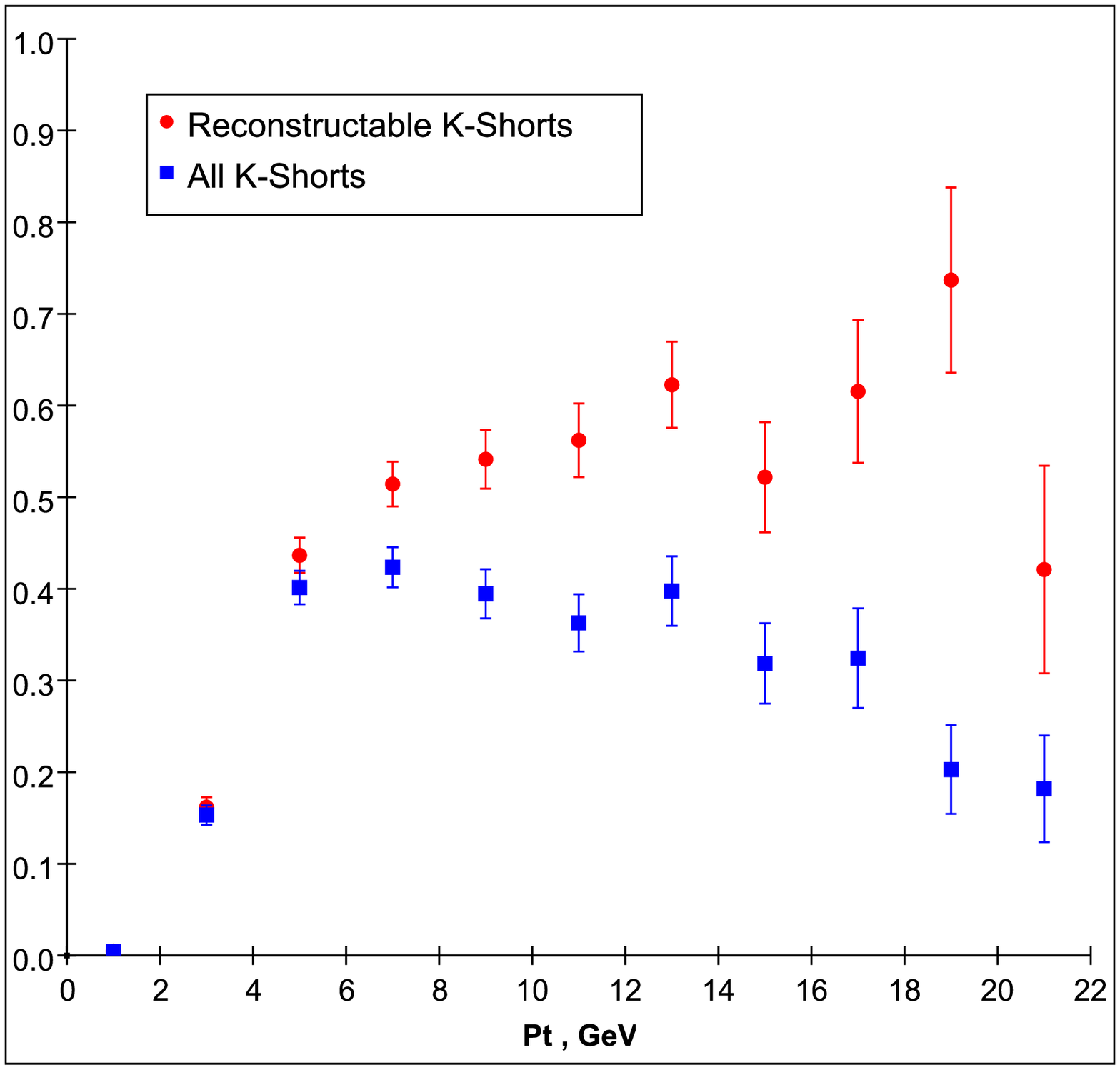}
\caption{$K_{S}^{0}$ reconstruction efficiency as a function of transverse momentum. 500~GeV $t\bar{t}$ events.}
\label{Fig:3}
\end{minipage}
\hspace{0.5cm} 
\begin{minipage}[t]{0.5\linewidth}
\centering
\includegraphics[width=1\linewidth]{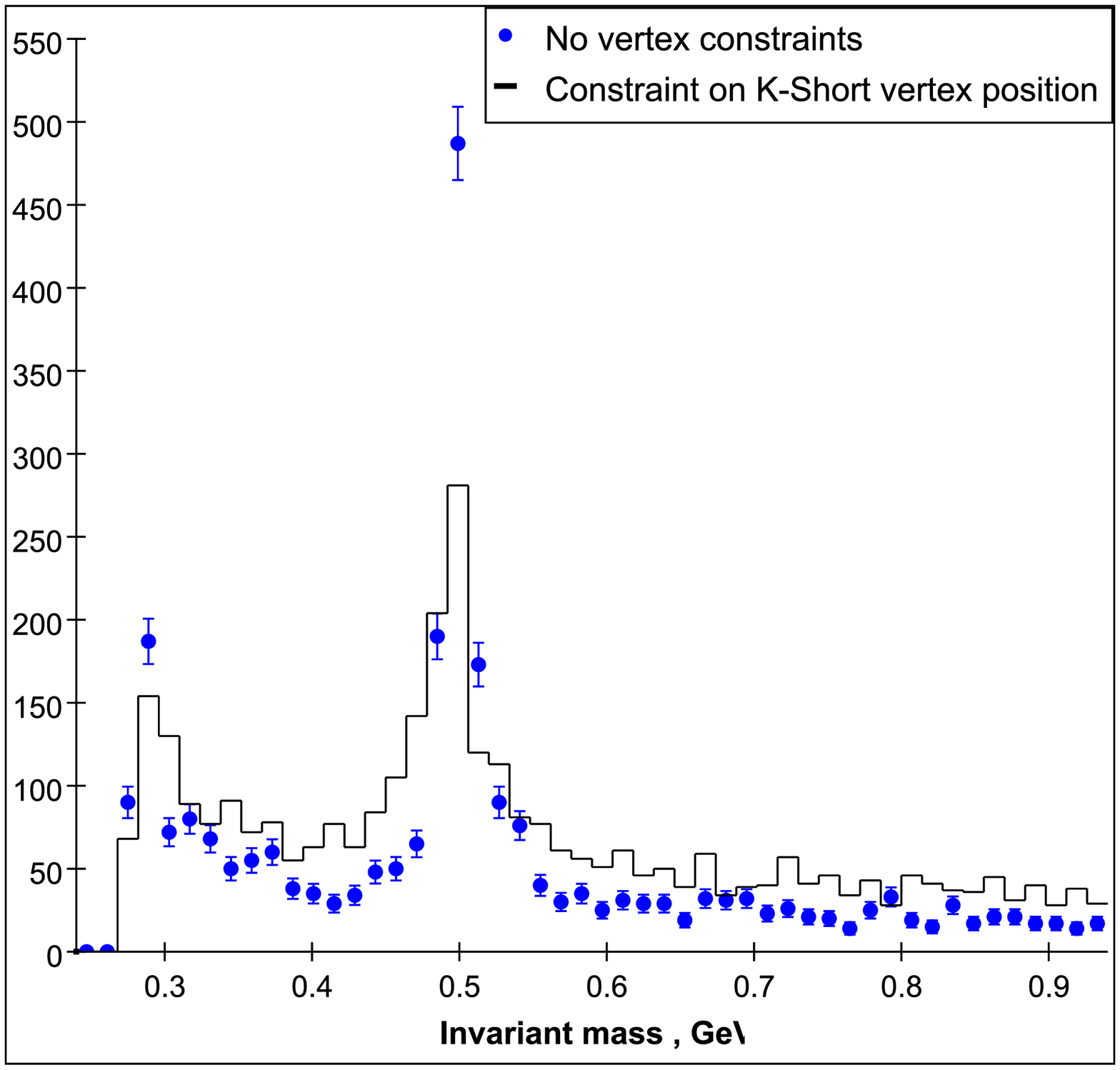}
\caption{Reconstructed $\pi^+\pi^-$ mass. 500~GeV $t\bar{t}$ events.}
\label{Fig:4}
\end{minipage}
\end{figure}

Figure \ref{Fig:3} shows $K_{S}^{0}$ reconstruction efficiency in $t\bar{t}$ events at
500~GeV center-of-mass energy. For comparison, less than 3~\% of all $K_{S}^{0}$
produced in such events would have been reconstructed by the vertex detector
seeded algorithm.  Figure \ref{Fig:4} shows reconstructed $K_{S}^{0}$
mass peak in $t\bar{t}$ events.

\section{Discussion}

The calorimeter assisted tracking algorithm addresses one of the critical issues for
the proposed SiD detector - reconstruction of long lived particles. It can also be 
instrumental in reconstructing kinked tracks that lose a substantial portion of
their energy in the tracker, as well as calorimeter backscatters. Availability of this
algorithm significantly improves overall robustness of the track reconstruction 
in SiD, reducing its reliance on the vertex detector.

The performance tests described in the previous section have been carried out
using the current version of the algorithm implementation. Several enhancements
are already in the works, and we expect substantial performance improvements.

One of the areas where improvements are desirable is reconstruction of low momentum
tracks. As seen in Figure \ref{Fig:1}, the reconstruction efficiency falls sharply
for charged pions with transverse momenta below 1 GeV. Many of these tracks never
reach the calorimeter barrel and, after leaving many hits in the tracker, enter 
calorimeter endcaps at shallow angles, making accurate determination of track
parameters from MIP stubs difficult. Our preliminary studies indicate, however,
that more flexible fitting and track extrapolation procedures may let us recover a
substantial portion of these tracks. We expect to implement these procedures once 
the infrastructure required to support them becomes available in the org.lcsim
framework.

The package implementing the calorimeter assisted tracking algorithm will remain
under continuing development in the near future. However, a fully functional 
version will be maintained in the org.lcsim production area. It will be
used for SiD geometry optimization and physics reach studies.


\begin{footnotesize}

\end{footnotesize}


\end{document}